\documentclass[nofootinbib,aps,preprint,pra,,showpacs]{revtex4-1}
\usepackage{graphicx}

\begin{document}

\title{What is nonlocal in counterfactual quantum communication?}

\author{Yakir Aharonov}
\affiliation{School of Physics and Astronomy, Tel Aviv University, Tel Aviv 69978, Israel\\ and Department of Physics, Chapman University, Orange CA, USA.}

\author{Daniel Rohrlich}
\affiliation{Department of Physics, Ben-Gurion University of the Negev, Beersheba
84105 Israel}

\date{\today}

\begin{abstract}
We revisit the ``counterfactual quantum communication" of Salih {\it et al.} \cite{cfc}, who claim that an observer ``Bob" can send one bit of information to a second observer ``Alice" without any physical particle traveling between them.  We show that a locally conserved, massless current---specifically, a current of modular angular momentum, $L_z$ mod $2\hbar$---carries the one bit of information.  We integrate the flux of $L_z$ mod $2\hbar$ from Bob to Alice and show that it equals one of the two eigenvalues of $L_z$ mod $2\hbar$, either 0 or $\hbar$, thus precisely accounting for the one bit of information he sends her. We previously \cite{Crete} obtained this result using weak values of $L_z$ mod $\hbar$; here we do not use weak values.

\end{abstract}


\maketitle

\vskip 1 in

H. Salih, Z.-H. Li, M. Al-Amri and M.S. Zubairy \cite{cfc} describe a remarkable effect, which they call ``counterfactual quantum communication" (CQC): transmission (across a ``transmission channel") of information from a sender to a receiver  ``without any physical particles traveling between them." Y. Cao {\it et al.} \cite{cfcx} and I. Alonso Calafell {\it et al.} \cite{calafell} have demonstrated this effect experimentally. For all our familiarity with quantum nonlocality, the effect is startling. It involves neither nonlocal quantum correlations (which anyway do not transmit information) nor the relative phase of the Aharonov-Bohm effect.  If any effect evokes Einstein's famous phrase ``spooky action at a distance", it is this one.  Yet we show below that counterfactual quantum communication does, after all, depend on a conserved local current crossing the ``transmission channel" between Alice and Bob; it is a current of {\it modular} \cite{mod} angular momentum $L_z$ mod $2\hbar$.  Consistent with the analysis of Salih {\it et al.} \cite{cfc}, the conserved current is massless.  Our demonstration of the conserved local current indicates that the effect is, after all, not spooky; it also highlights the importance of modular variables in understanding quantum nonlocality.

We will describe a thought experiment equivalent to the one of Salih {\it et al.} \cite{cfc}.  But for clarity we begin, like \cite{cfc}, with a toy version of the experiment.  Figure 1 shows a particle wave packet in a cavity of length $L$, with Alice at the left end of the cavity (which is closed 
and reflects the particle), and Bob at the right end (which is closed and reflects the particle, but which Bob can open).  Halfway between the two ends is a thin barrier; it transmits the particle with (a small) amplitude $i\sin \epsilon$ and reflects it with amplitude $\cos \epsilon$.  Let the particle, with $\Delta x\ll L$ (as in Fig. 1) and a large expectation value $p$ of the momentum (such that $\Delta p \ll |p|$), start from Alice's end and hit the
barrier; it then either returns with amplitude $\cos \epsilon$, or continues towards Bob with amplitude $i\sin\epsilon$.  We can represent the evolution of these wave packets via a unitary matrix $U$:
\begin{equation}
\label{1}
U(\epsilon) = \pmatrix{ \cos \epsilon &
i\sin \epsilon \cr
i\sin \epsilon &
\cos \epsilon \cr} ~~~~.
\end{equation}
\medskip
Note that $U(j\epsilon) U(\epsilon)  = U([j+1]\epsilon)$ and, by induction, $[U(\epsilon)]^j = U(j\epsilon)$.  (Factors of $-1$ due to wave-packet bounces at the ends cancel.) Thus if the initial state of the particle is $\pmatrix{1\cr 0 \cr}$ (the particle is on Alice's side), then after $j$ laps back and forth to the barrier the particle is in the state $\pmatrix{\cos j\epsilon \cr i\sin j\epsilon\cr}$; and if $j\epsilon =\pi /2$, the particle is certain to be on Bob's side of the barrier.  Let $T$ denote the time required for the particle to get to Bob's side with certainty, and let $v$ denote the speed of the particle in either direction.  Then $j$ times back and forth correspond to a distance $jL$ and a time $jL/v$; taking $j =\pi/2\epsilon$, we obtain $T =\pi L/2\epsilon v$ as the time when the particle is on Bob's side with certainty.  At time $T$ the particle is on Bob's side, at time $2T$ it is back on Alice's side with an overall phase factor $-1 =\cos \pi$, etc.

\begin{figure}
\begin{center}
\includegraphics[width=15.cm, height=10.cm]{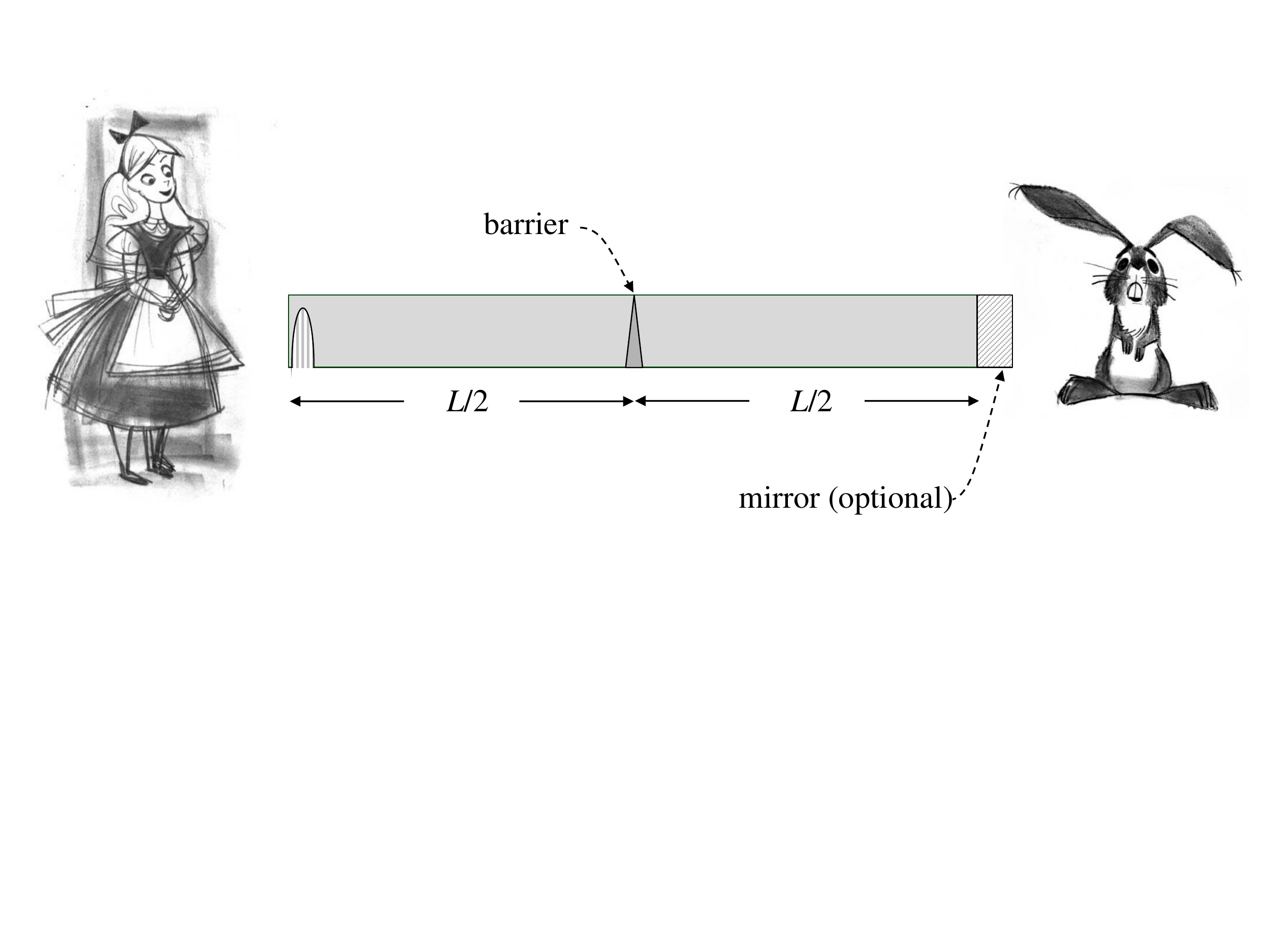}
\end{center}
\vskip -5 cm
\caption{Alice and Bob at opposite ends of a single cavity, with a barrier placed symmetrically between them.  A particle wave packet is initially at Alice's end.}
\label{fig1}
\end{figure}

We now consider a protocol that allows Bob to send a single bit of information to Alice. Bob has the option of keeping his end of the cavity closed with a mirror.  Every wave packet that arrives at his end is reflected.  Wave packets from Alice and Bob hit the barrier simultaneously, interfering constructively on Bob's side, and at time $T$, after $j$ laps such that $j\epsilon = \pi/2$, the particle is certainly at Bob's end. Bob's other option is to remove the mirror and leave his end of the cavity open.  Now with probability $\sin^2 \epsilon$, a wave packet that hits the barrier will continue on to Bob and escape the cavity altogether. Ultimately, so would the particle.  But ``ultimately" corresponds to unlimited wave packet hits on the barrier, i.e. to an arbitrarily long experiment.  As long as the time is finite (as it must be in any realistic experiment), the number of hits is finite (even if proportional to $1/\epsilon$) and $\sin^2\epsilon$ times that number is arbitrarily small, i.e. nothing enters the transmission channel. In effect Bob, by removing his mirror, turns the $barrier$ into a mirror, such that the particle cannot leave Alice's side of the cavity.  Then Alice, by checking her end of the cavity at time $T$, learns Bob's choice:  if she finds the particle there, Bob removed the mirror; if not, he left the mirror covering his end.  Bob sends Alice one bit of information and, if he removed the mirror, no physical particle traveled between Alice and Bob, to order $\epsilon$.  Still, if Bob left the mirror covering his end, the particle did travel from Alice to Bob (and back).  Thus Alice and Bob have not achieved $complete$ counterfactual quantum communication.

Note, we have avoided writing $\epsilon \rightarrow 0$ because it suggests the unphysical limit $T\rightarrow \infty$. A physical experiment cannot last forever, hence $T$ must be finite (though $T$ can be arbitrarily large).  Below we define $\epsilon_A$, $\epsilon_B$ and $\epsilon_B/ \epsilon_A$ and their limits $\epsilon_A\rightarrow 0$, $\epsilon_B\rightarrow 0$ and $\epsilon_B/\epsilon_A \rightarrow 0$ such that $\epsilon_A$, $\epsilon_B$ and $\epsilon_B/\epsilon_A$ are arbitrarily small but $\epsilon_A \ne 0 \ne \epsilon_B$.

For complete CQC, consider Fig. 2.  Near Bob's end, there are two thin barriers.  Barrier $A$ transmits with amplitude $i\sin \epsilon_A$ and reflects with amplitude $\cos \epsilon_A$; barrier $B$ transmits with amplitude $i\sin \epsilon_B$ and reflects with amplitude $\cos \epsilon_B$.  The distance between Alice's end and barrier $A$ is still $L/2$, and we set it equal to $n_B$ times the distance between barriers $A$ and $B$. Thus, in the time $L/v$ it takes a wave packet to bounce from barrier $A$ to Alice's end and then back to barrier $A$, a wave packet between barriers $A$ and $B$ could bounce back and forth $n_B$ times.  Now let $n_B\epsilon_B =\pi/2$.  Then a wave packet bouncing between barriers $A$ and $B$ can escape through $B$ in the time $L/v$ it takes a wave packet to reflect to Alice from barrier $A$ and then bounce back to barrier $A$.

And now we add a restriction: barrier $A$ is completely closed off (in effect, $\epsilon_A =0$) $except$ at times $t = L/2v, 3L/2v, 5L/2v,\dots$.  Thus if Alice releases a wave packet from her end at time $t=0$, it reaches barrier $A$ at time $t=L/2v$, and passes through barrier $A$ with amplitude $i \sin \epsilon_A$ or reflects from it with amplitude $\cos \epsilon_A$.  If the particle wave packet arrives at barrier $A$ at any time $t$ that is not on the list $t = L/2v, 3L/2v, 5L/2v,\dots$, it reflects with amplitude 1. Note, this restriction on barrier $A$ simplifies the evolution, as follows:  between two consecutive approaches of the wave packet to barrier $A$ from Alice's end (e.g. between times $t=L/2v$ and $t=3L/2v$), the wave packet bouncing between barriers $A$ and $B$ cannot pass through barrier $A$, because barrier $A$ is completely closed off. Thus without loss of generality we can take $t$ from the list (because otherwise the evolution is trivial) and even take the initial time to be $t=0$.

\begin{figure}
\begin{center}
\includegraphics[width=15.cm, height=10.cm]{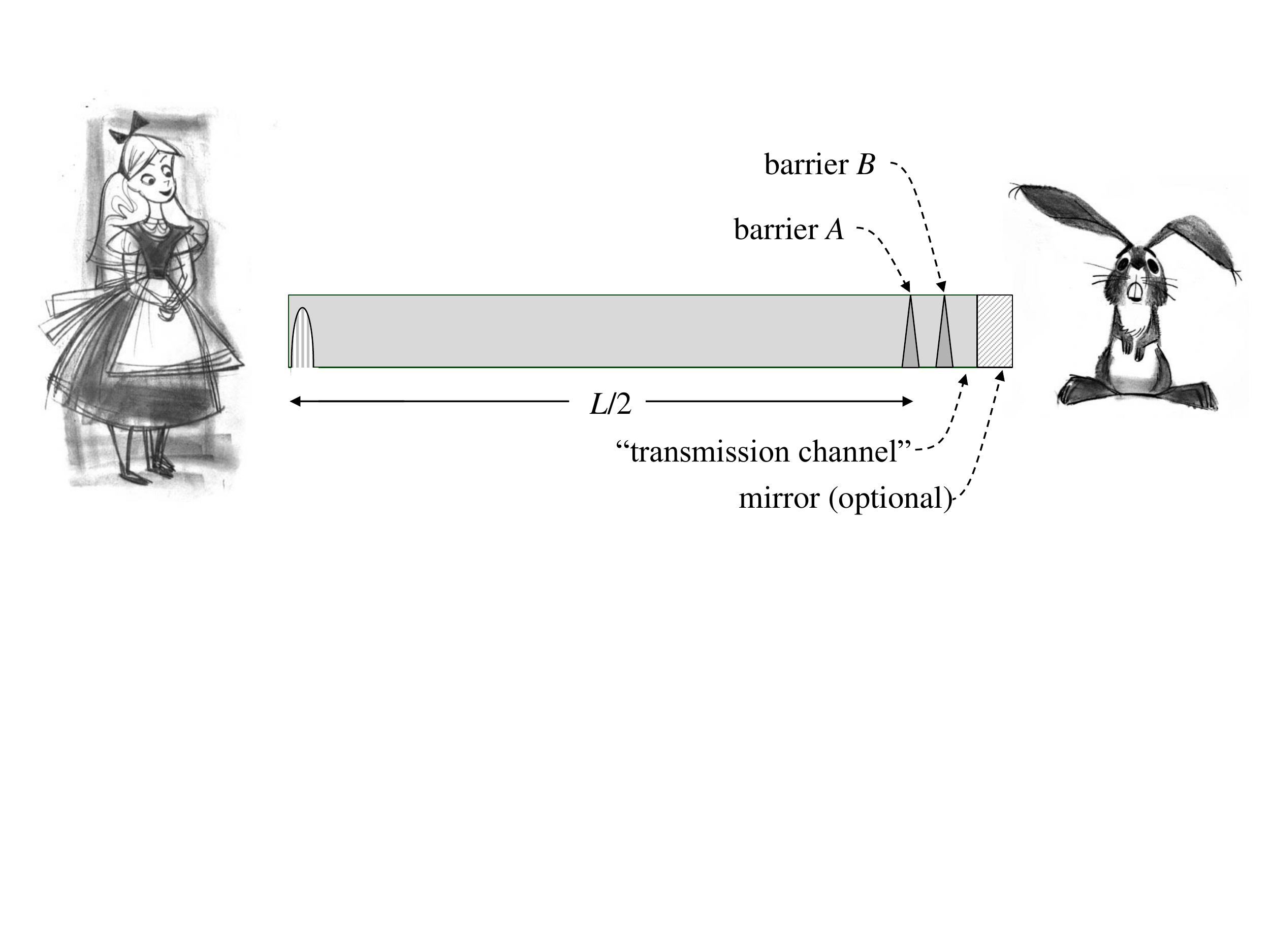}
\end{center}
\vskip -5 cm
\caption{Alice and Bob at opposite ends of an asymmetrical cavity, with two barriers between them.  A particle wave packet is initially at Alice's end.
}
\label{fig2}
\end{figure}

Let us now consider all possible evolutions of Alice's wave packet once it reaches barrier $A$.  Let $j_A$ index the collisions (separated in time by $L/v$) of the wave packet on Alice's side with barrier $A$; we take $1\le j_A \le n_A$, defining $n_A$ implicitly via $n_A\epsilon_A =\pi/2$ in analogy with $n_B$.  Likewise, we let $j_B$ index the collisions with barrier $B$ of a wave packet between the barriers. We define a protocol for Bob to choose ``logic 0" or ``logic 1": Bob chooses ``logic 0" by $not$ closing his end of the cavity (i.e. ``0" indicates ``zero closing").  What is the probability that a wave packet crossing from Alice's side will exit to the right (through barrier $B$)?  We ask this question not only to understand the effect of Bob's choice, but also to keep the protocol counterfactual.  Once the wave packet crosses barrier $A$ in the direction of barrier $B$, it bounces up to $n_B$ times between the barriers; and for each bounce, the probability that it escapes to the right is $\sin^2 \epsilon_B$.  The total probability is then $n_B \sin^2\epsilon_B$, which is negligible for $\epsilon_B\rightarrow 0$. The particle does not pass beyond barrier $B$; it reaches barrier $B$ and crosses barrier $A$ back to Alice. As in the toy version with Bob's mirror in place, it spends some time on Alice's side of barrier $A$ and some time on Bob's side, but never crosses barrier $B$.

Now, for the $j_A$-th collision from Alice's side, the probabilities for the particle to reflect back to her or cross over to Bob are $\cos^2 (j_A \epsilon_A)$ and $\sin^2(j_A\epsilon_A)$, respectively, in analogy with Eq. (\ref{1}).  This result now tells us how to keep the protocol counterfactual.  The total probability for the particle to exit right for $1\le j_A \le n_A$ is $\sum_{j_A =1}^{n_A}  \sin^2 (j_A\epsilon_A) n_B (\epsilon_B)^2$, which we can rewrite as $(n_A/2) n_B (\epsilon_B)^2$, since the average value of $\sin^2(j_A\epsilon_A)$ in the range $1\le j_A \le n_A$ is $1/2$. Hence the condition $\epsilon_A \rightarrow 0$ does not by itself keep this protocol counterfactual, i.e. does not zero the probability that the particle enters the transmission channel.  We must require $(n_A/2) n_B (\epsilon_B)^2 = (\pi^2/8) \epsilon_B/\epsilon_A \rightarrow 0$ as well, i.e. both $\epsilon_A \rightarrow 0$ and $\epsilon_B/\epsilon_A \rightarrow 0$.  We will refer to this stricter condition as the ``double limit". Thus for ``logic 0" and in the double limit, no part of the wave function exits right. At time $T_A=\pi L /2\epsilon_A v$ (after $n_A =\pi /2\epsilon_A$ trips between Alice and barrier $A$, each taking time $L/v$), the particle is on Bob's side of barrier $A$; at time $2T_A$ it is back to Alice with an overall phase factor $-1$; and so on.

For ``logic 1", Bob closes his end of the cavity with a perfectly reflecting mirror.  At $j_A=1$, a single wave packet of amplitude $i\sin \epsilon_A$ passes through barrier $A$, then self-interferes through barrier $B$ into a single wave packet with amplitude $i\sin \epsilon_A$ that is about to hit Bob's mirror. Bob briefly removes the mirror, and nothing reflects left.  Yet also the probability of finding the particle to the right of barrier $B$ is negligible for $\epsilon_A\rightarrow 0$.  For successive values of $j_A$ the evolution is the same, except that a factor $\cos^{j_A -1} (\epsilon_A)$ multiplies the amplitude; but for $\epsilon_A \rightarrow 0$, it reduces to 1.  Thus for ``logic 1", Alice always finds the particle on her side---not because it returned but because it never left, as in the toy version without Bob's mirror.  Indeed, the probability for Alice to find the particle on her side after $n_A$ reflections approaches 1 for $\epsilon_A \rightarrow 0$.

Have we achieved counterfactual quantum communication with this protocol?  Indeed, we have. Bob is beyond the cavity, at its very end, which he either covers or doesn't cover with a mirror.  He is never in the ``transmission channel" between barrier $B$ and his end of the cavity.  Nor is the particle ever in the transmission channel, for $\epsilon_A\rightarrow 0$ and $\epsilon_B/\epsilon_A \rightarrow 0$.  For logic 1, the total amplitude to be in the transmission channel cannot be greater than $\epsilon_A$.  After $n_A$ times that the wave packet on Alice's side hits barrier $A$, the total probability that the particle enters the transmission channel is at most $n_A |\epsilon_A|^2 = (\pi /2\epsilon_A) \vert\epsilon_A \vert^2=\pi \epsilon_A/2$ and is negligible for $\epsilon_A \rightarrow 0$ (as in the case of logic 0). Bob uses the particle to send one bit to Alice: if the wave packet is consistently at her end of the cavity then Bob chose ``logic 1"; if the wave packet is at her end only at specified times, then Bob chose ``logic 0".  Either way, the particle never crosses the transmission channel separating them.

So far we have merely argued, with Salih {\it et al.} \cite{cfc}, that complete counterfactual quantum communication is possible.  We now show, however, that this argument leads to a paradox:  it violates Noether's fundamental theorem on symmetries and conservation laws.  Figure 3 shows a variation on our experiment, with a $z$ axis and two identical cavities placed parallel to it, symmetrically above and below. Define $L_z \equiv -i\hbar\partial /\partial \phi$ where $\phi$ is the conjugate angle about the $z$ axis; then $e^{i\pi L_z/\hbar}$ effects a $\pi$ rotation about the $z$ axis. If $\vert\psi_\uparrow\rangle$ and $\vert\psi_\downarrow\rangle$ represent the same state but in the upper and lower cavities, respectively, then the operator $e^{i\pi L_z/\hbar}$ interchanges them (up to a relative phase).

Now suppose Bob applies logic 0 to the upper cavity and logic 1 to the lower cavity.  Figure 3 shows the wave packet in a symmetric superposition above and below the $z$ axis (at Alice's end).  The cavities in which the particle moves (in a superposition) are invariant under a rotation of $\pi$ about the $z$ axis, and the operation $e^{i\pi L_z/\hbar}$ leaves the state of the particle invariant.  And herein lies the paradox.  We have seen that the particle's evolution depends sensitively on the conditions in the transmission channel---whether or not there is a mirror, etc.  But we have also seen that in the double limit $\epsilon_A \rightarrow 0$ and $\epsilon_B/\epsilon_A \rightarrow 0$, the probability that the particle enters the transmission channel is negligible!  If the particle doesn't enter the transmission channel, then the potential {\it in which it moves} is invariant under a $\pi$ rotation $e^{i\pi L_z/\hbar}$ and the ``modular angular momentum" $L_z$ mod $2\hbar$ must be conserved.  (Note that, in the exponent, $L_z$ is automatically $L_z$ mod $2\hbar$.) But, it seems, modular angular momentum is $not$ conserved: if the initial phase between the components of the particle is set to 0, after a time $2T_A$ it will be $\pi$, and vice versa, corresponding to a shift in angular momentum (mod $2\hbar$) of $\hbar$.  This paradox arises when we consider the particle in a superposition of states in two cavities.  A related paradox (Fig. 4) has just one cavity, with Bob's $mirror$ in a superposition that is invariant under a $\pi$ rotation about the $Z$ axis.  The particle in the cavity, facing a superposition of logics 0 and 1, becomes entangled with the position of the mirror. After a time $2T_A$, the particle and mirror disentangle, with only a relative phase factor $-1$ between the mirror positions testifying to the transient entanglement.  What happened to conservation of modular angular momentum?  Do the particle and the mirror exchange $L_z$? Both paradoxes suggest that the particle entered a region from which it was excluded.

\begin{figure}
\begin{center}
\includegraphics[width=15.cm, height=10.cm]{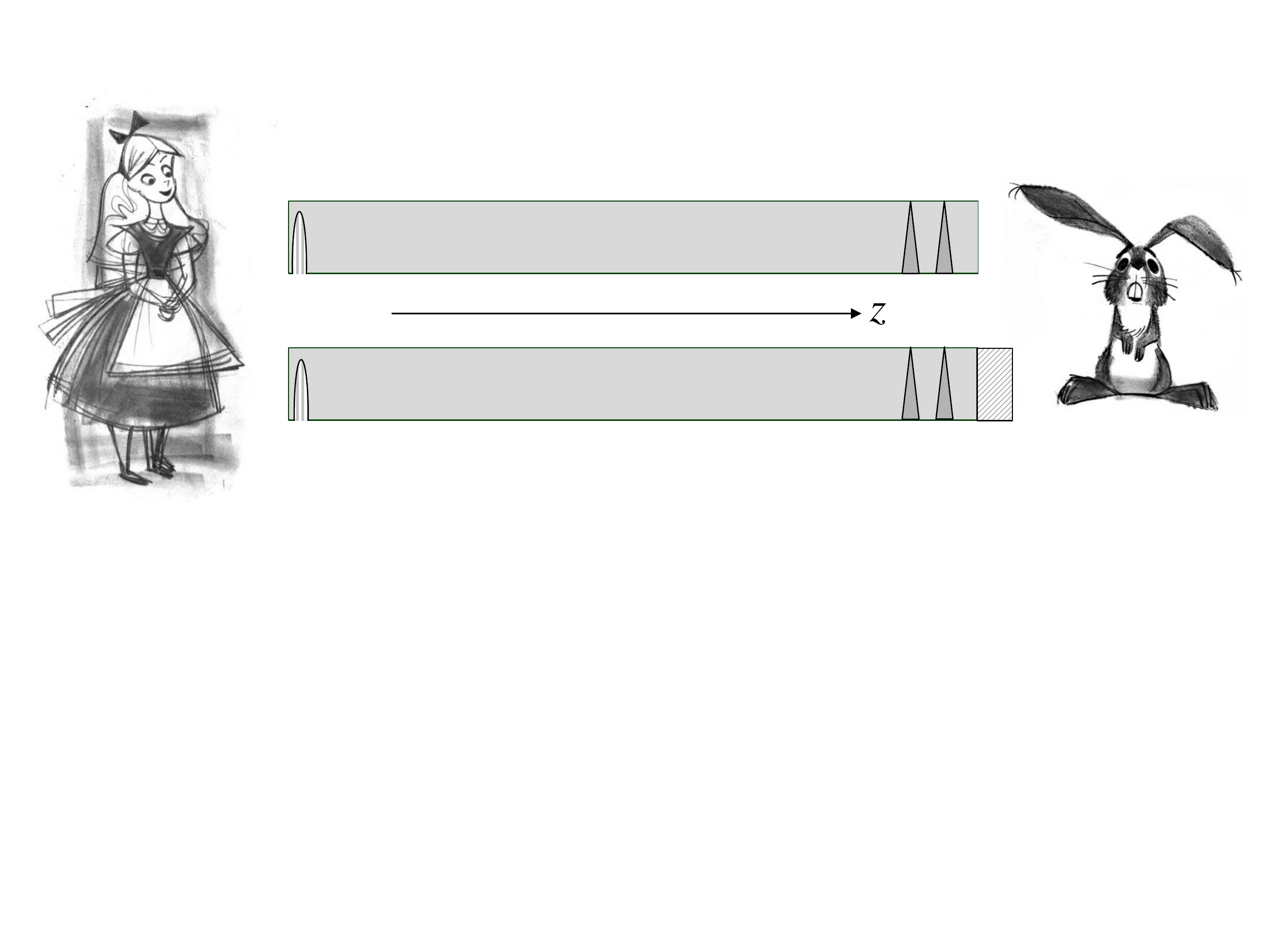}
\end{center}
\vskip -5 cm
\caption{Alice and Bob at opposite ends of $two$ cavities, symmetrically about the $z$ axis.  The potential (excluding the mirror) is invariant under a $\pi$ rotation about the $z$ axis, as is the initial particle superposition.
}
\label{fig3}
\end{figure}

As Sherlock Holmes would say, ``When you have eliminated the impossible, whatever remains, however improbable, must be the truth!"  We have, with Salih {\it et al.} \cite{cfc}, eliminated the possibility that the particle entered the transmission channel; therefore, its evolution {\it must conserve} modular angular momentum, however implausible that might seem.

What we now calculate is the $flow$ of modular angular momentum across barriers $A$ and $B$ into the transmission channel, for the case of the first paradox---the two parallel cavities of Fig. 3.  We do this by calculating the expectation value of $e^{i\pi L_z/\hbar}$ in a state that is an initial superposition $\left[ \vert \psi_\uparrow\rangle +\vert \psi_\downarrow \rangle \right]/\sqrt{2}$, where $\psi_\uparrow$ and $\psi_\downarrow$ indicate localized particle amplitudes in the upper and lower cavities in Fig. 3, respectively; $\psi_\uparrow$ and $\psi_\downarrow$ depend implicitly on $j_A$ and $j_B$, and thus on time.  To the lower cavity Bob applies logic 1, and then nothing from Alice's end of the lower cavity tunnels through barrier $A$.  The amplitude on Alice's side is $\cos^{j_A} (\epsilon_A)$, which is simply 1 in our double limit $\epsilon_A \rightarrow 0$ and $\epsilon_B/\epsilon_A \rightarrow 0$:  the particle in the lower cavity never leaves Alice.   At the same time, to the upper cavity Bob applies logic 0; then any wave packet crossing barrier $B$ never returns, but tunnels into the transmission channel with vanishing probability.  Thus the probability that the particle remains on Alice's side is $\cos^2 (j_A \epsilon_A)$ while the probability that it tunnels to (and reflects from) barrier $B$ is $\sin^2 (j_A \epsilon_A)$, as we saw.

We have accounted for the probabilities in detail.  But we have not yet calculated the expectation value $\langle e^{i\pi L_z/\hbar} \rangle$.  Since $e^{i\pi L_z/\hbar}$ interchanges $|\psi_\uparrow \rangle$ and $|\psi_\downarrow\rangle$, we have $\langle e^{i\pi L_z/\hbar} \rangle =  \Re \langle \psi_\uparrow \vert \psi_\downarrow\rangle$, where $\Re$ indicates the real part.  Thus $\langle e^{i\pi L_z/\hbar} \rangle$ contains a contribution $\cos (j_A \epsilon_A)$ from Alice's side of barrier $A$, but none from between barriers $A$ and $B$ since, in the lower cavity, the particle never enters the region
between barriers $A$ and $B$.  (See Table 1.)  Does the transmission channel contribute to $\langle e^{i\pi L_z/\hbar} \rangle$?
\begin{figure}
\begin{center}
\includegraphics[width=15.cm, height=10.cm]{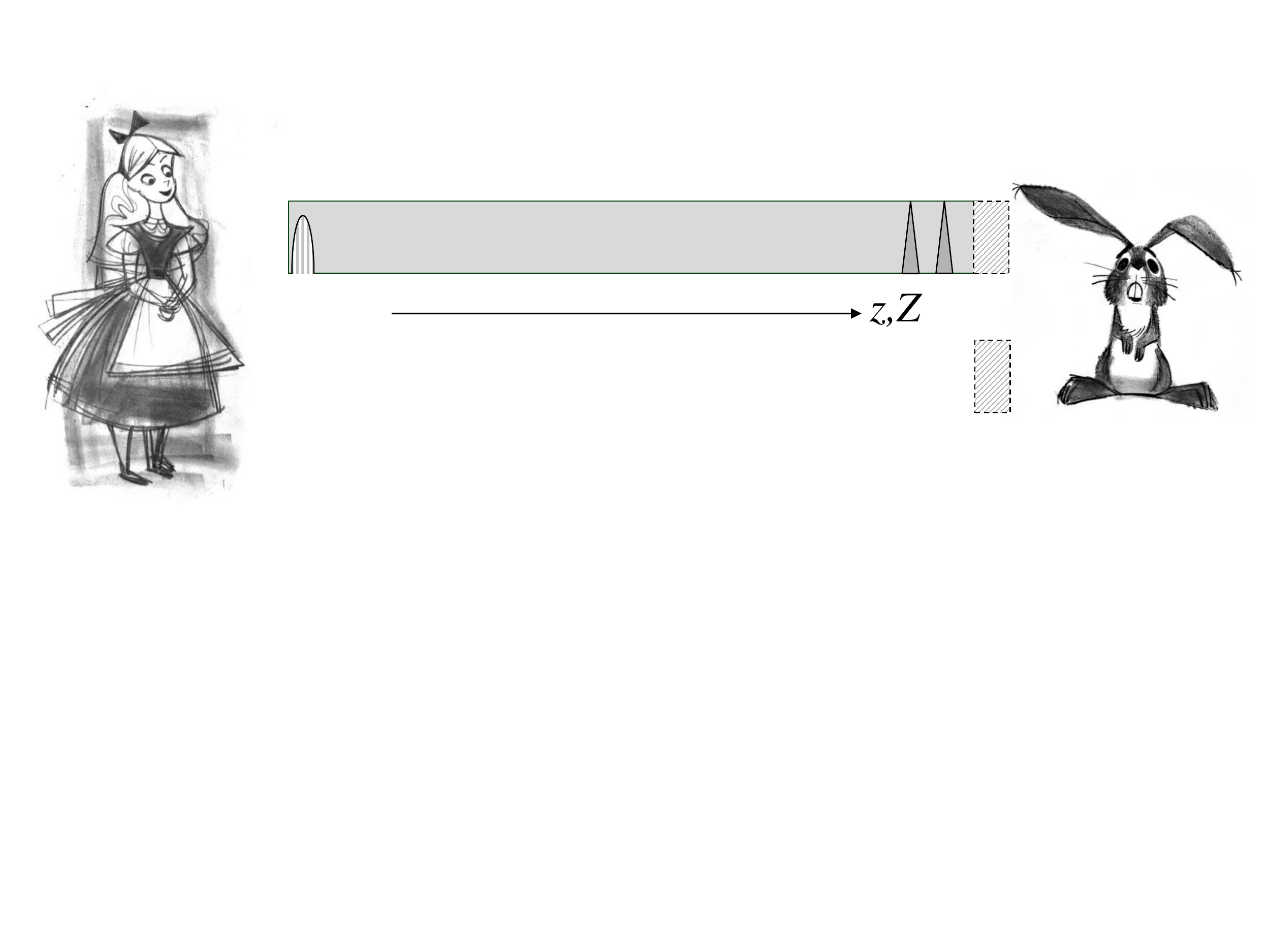}
\end{center}
\vskip -5 cm
\caption{Alice and Bob again share a single cavity, with the mirror at Bob's end in a superposition of two orthogonal states.
}
\label{fig4}
\end{figure}

The amplitude for the particle to enter the transmission channel is a product of factors that depend on Bob's choice of logic 1 or logic 0.  First, logic 0:  the amplitude for the particle to pass through barrier $A$ is $i\sin (j_A \epsilon_A)$, as calculated.  Once it passes through barrier $A$, there is an amplitude $\cos^{j_B -1}(\epsilon_B) (i\sin\epsilon_B)$ for the particle to cross barrier $B$ into the transmission channel on the $(j_B-1)$-th collision with it.  The product amplitude is then $i\sin (j_A \epsilon_A) \cos^{j_B -1} (\epsilon_B) (i\sin\epsilon_B)$, which we can replace with $-\epsilon_B \sin (j_A \epsilon_A)$ in our double limit. Next, logic 1: the amplitude for the particle to pass through barrier $A$ is $\cos^{j_A - 1} (\epsilon_A) (i\sin \epsilon_A)$, since the particle never returns to Alice's side and only reflects $j_A - 1$ times with amplitude $(\cos \epsilon_A)^{j_A - 1}$ before it finally passes through on the $j_A$-th collision, with amplitude $\sin \epsilon_A$.  It then enters between barriers $A$ and $B$ where (since Bob has covered his end with a mirror) its amplitude to reach the transmission channel is $\sin (j_B \epsilon_B)$. Multiplying the amplitudes, we get $\cos^{j_A - 1} (\epsilon_A) (i\sin \epsilon_A) i\sin (j_B \epsilon_B)$. But in our double limit, we can replace $\cos^{j_A - 1} (\epsilon_A)$ by 1 and $\sin \epsilon_A$ by $\epsilon_A$, to get $-\epsilon_A \sin (j_B \epsilon_B)$ as the amplitude in the transmission channel for logic 1. Now for the product of products in the last column we obtain $\epsilon_A \sin (j_B \epsilon_B) \epsilon_B \sin (j_A \epsilon_A)$, which factors into $\left[ \epsilon_A \sin (j_A\epsilon_A) \right] \times \left[ \epsilon_B \sin (j_B\epsilon_B) \right]$, as the contribution to $\langle e^{i\pi L_z/\hbar}\rangle$ from the transmission channel, for a given $j_A, j_B$.

\begin{table}
    \begin{tabular}{
     p{3cm} | p{2cm} | p{5.5cm} | p{5.5cm} |}
    {}& {Alice's side} & {Between $A$ and $B$} & {Transmission channel}\\ \hline
    logic 0 (open) & $\cos (j_A\epsilon_A)$ & $i\sin(j_A\epsilon_A)\cos^{j_B}(\epsilon_B)$

$\rightarrow i\sin (j_A\epsilon_A)$ & $i\sin(j_A\epsilon_A)\cos^{j_B-1}(\epsilon_B) (i\sin\epsilon_B)$
    $\rightarrow -\epsilon_B \sin(j_A\epsilon_A)$ \\ \hline
    logic 1 (covered)  & $\cos^{j_A} (\epsilon_A)$
    $\rightarrow 1$& $\cos^{j_A-1} (\epsilon_A)(i\sin \epsilon_A ) \cos (j_B\epsilon_B )$
    $\rightarrow i\epsilon_A \cos(j_B \epsilon_B )$ & $\cos^{j_A-1}(\epsilon_A) (i\sin \epsilon_A) i\sin (j_B\epsilon_B )$

    $\rightarrow -\epsilon_A \sin (j_B \epsilon_B)$ \\
    \hline
    \end{tabular}
    \medskip
\caption{Particle amplitudes in three regions as functions of $j_A$, $j_B$ and ``logic 0" (top row) or ``logic 1" (bottom row).  The arrows indicate how the amplitudes simplify towards the ``double limit".  In each row, the squared absolute values of the amplitudes sum to 1.}
\end{table}

We now sum over $j_A$ and $j_B$.  Why? Consider an evolution in which the particle, initially on Alice's side, passes through barrier $A$ and reflects between barriers $A$ and $B$ a total of $n_B$ times, ultimately crossing back to Alice.  Although there are $n_B$ reflections off barrier $B$, there is only one ``path", in the sense of Feynman's sum over paths.  By contrast, the evolution in the transmission channel is a sum over ``paths" in which each pair $j_A ,j_B$ defines a $unique$ path with the particle crossing barrier $A$ on the $j_A$-th collision with it and entering the transmission channel on the $j_B$-th collision with barrier $B$.  There are $j_A j_B$ distinct paths.  We first sum over $j_B$ from 1 to $n_B$. Since $\epsilon_B = \pi /2n_B$, the sum simplifies to an integral in our double limit:
\begin{equation}
\label{sum}
\lim_{\epsilon_B \rightarrow 0} \sum_{j_B=1}^{n_B} \epsilon_B \sin{j_B \epsilon_B}
= \lim_{n_B \rightarrow \infty}\sum_{j_B=1}^{n_B} {\pi \over {2n_B}} \sin \left( {j_B\pi / {2n_B}}\right)
=  {\pi \over 2}\int_0^1 dx  \sin (\pi x/2) =1~~.
\end{equation}
Then replacing $A$ for $B$ everywhere in Eq. (\ref{sum}), we get the same result, namely 1.

But instead of setting $n_A$ equal to $\pi /2\epsilon_A$, as we did for Bob's protocol, we can leave it arbitrary.  Then the corresponding integral in Eq. (\ref{sum}) approaches (in the double limit) $1-\cos(n_A\epsilon_A)$.  Now let us sum the contributions to $\langle e^{i\pi L_z/\hbar}\rangle$ from the entire cavity.  The entire cavity includes Alice's side, which contributes $\cos (n_A\epsilon_A)$ to $\langle e^{i\pi L_z/\hbar}\rangle$; the intermediate region between barriers $A$ and $B$, which contributes nothing; and the transmission channel, which contributes $1-\cos (n_A\epsilon_A)$.  For any value of $n_A$, the sum equals 1.  There is no loss of modular angular momentum; the total $L_z$ mod $2\hbar$ (summed over the range of $z$) remains a constant of the motion.  And, though we sum over $j_A$ and $j_B$, each pair $j_A$, $j_B$ corresponds to a $time$ in the evolution of the particle wave packet---the time at which Alice and Bob have completed their respective numbers of laps. The flux of modular angular momentum across barriers $A$ and $B$ into the transmission channel changes with time.

We thus arrive at a striking resolution of the paradox of Fig. 3:  the expectation value $\langle e^{i\pi L_z/\hbar}\rangle =1$ is invariant after all, but the particle never enters the transmission channel!  What enters the transmission channel is the angular momentum of the particle, mod $2\hbar$.  We have integrated the flux of $L_z$ mod $2\hbar$ between Bob and Alice and shown that it equals one of the two eigenvalues of $L_z$ mod $2\hbar$, namely 0 or $\hbar$, accounting for the one bit of information he sends her.  Modular angular momentum is a nonlocal dynamical variable, but the particle and its modular angular momentum separate $locally$---at a point---without action at a distance (just as the ``weak Cheshire cat" parts ways locally with its grin \cite{current}).  In the paradox of Fig. 4, the particle itself never entered the space of the mirror, but $\hbar$ mod $2\hbar$ of its modular angular momentum flowed into it, and would show up as an $\hbar$ shift in $L_Z$ mod $2\hbar$, i.e. as a $\pi$ rotation of the mirror wave function.

Note that we can demonstrate $local$ conservation of the $nonlocal$ quantities $\langle e^{i\pi L_z/\hbar} \rangle$ and $\langle e^{i\pi L_Z/\hbar} \rangle$ only because $z$ and $Z$, the locations of the particle and mirror along the symmetry axis, commute with these conserved quantities. With respect to locality, modular angular momentum is a $hybrid$ quantity: $\langle e^{i\pi L_z/\hbar} \rangle$ and $\langle e^{i\pi L_Z/\hbar} \rangle$ are locally conserved along the $z,Z$ axis although they act nonlocally in the perpendicular plane (e.g. reveal relative phases between the upper and lower wave packets in Fig. 3). 
There is no modular angular momentum in Fig. 2, but there is modular energy $H$ mod $h/2T$, and Bob's choice of logic 0 or logic 1 governs the distribution of modular energies at Alice's end. Our calculation is not a full treatment of quantum conserved currents, yet we have demonstrated a new approach to nonlocality in quantum mechanics.  Quantum mechanics is indeed nonlocal.  But instead of ``spooky action at a distance", we uncover here a much more satisfactory interpretation of counterfactual quantum communication:  properties of a particle can travel $locally$ through regions from which the particle itself is excluded.  Quantum mechanics is not, after all, so spooky \cite{spooky}.
\goodbreak

\begin{acknowledgments}
We sincerely thank Referee A for astutely grasping the novelty of our paper, and all the Referees for their clear insight into what we had left unclear. Y.A. thanks the Israel Science Foundation (Grant No. 1311/14), the ICORE Excellence Center ``Circle of Light", and the German-Israeli Project Cooperation (DIP) for support. D.R. thanks the Israel Science Foundation (Grant No. 1190/13) and the John Templeton Foundation (Project ID 43297) for support.  The opinions expressed in this publication are the authors' and do not necessarily reflect the views of any of these supporting foundations.
\end{acknowledgments}

\goodbreak


\begin{references}
\bibitem{spooky} At least, not in the sense of explicit action at a distance.
\end{references}
\end{document}